\documentclass[runningheads]{llncs}
\usepackage[T1]{fontenc}
\usepackage{graphicx}
\usepackage{xcolor}
\usepackage[linesnumbered,ruled,vlined]{algorithm2e}
\usepackage{hyperref}
\usepackage{amsmath}

\begin{document}
\title{High-Throughput LLM inference on Heterogeneous Clusters}
%
%
\author{Yi Xiong\inst{1} \and
Jinqi Huang\inst{2}\orcidID{0000-0003-3913-0067} \and
Wenjie Huang\inst{2} \and
Xuebing Yu\inst{2} \and
Entong Li\inst{2} \and
Zhixiong Ning\inst{2} \and
Jinhua Zhou\inst{2} \and
Li Zeng\inst{2}\orcidID{0000-0002-6537-4084} \and
Xin Chen\inst{2}} 
\authorrunning{Y. Xiong et al.}
%
\institute{Huawei Technologies Co., Ltd\\
\email{\{xiongyi26,huangjinqi1,huangwenjie16,yuxuebing3,lientong,\\
	ningzhixiong1,zhoujinhua1,zengli43,chenxin\}@huawei.com}}
\maketitle              
\begin{abstract}
Nowadays, many companies possess various types of AI accelerators, forming heterogeneous clusters. Efficiently leveraging these clusters for high-throughput large language model (LLM) inference services can significantly reduce costs and expedite task processing.
However, LLM inference on heterogeneous clusters presents two main challenges. Firstly, different deployment configurations can result in vastly different performance. 
The number of possible configurations is large, and evaluating the effectiveness of a specific setup is complex. 
Thus, finding an optimal configuration is not an easy task. 
Secondly, LLM inference instances within a heterogeneous cluster possess varying processing capacities, leading to different processing speeds for handling inference requests. 
Evaluating these capacities and designing a request scheduling algorithm that fully maximizes the potential of each instance is challenging.
In this paper, we propose a high-throughput inference service system on heterogeneous clusters. 
First, the deployment configuration is optimized by modeling the resource amount and expected throughput and using the exhaustive search method. 
Second, a novel mechanism is proposed to schedule requests among instances, which fully considers the different processing capabilities of various instances.
Extensive experiments show that the proposed scheduler improves throughput by 122.5\% and 33.6\% on two heterogeneous clusters, respectively.

\end{abstract}
\section{Introduction}
In recent years, large language models (LLMs) such as GPT-4 \cite{openai2024gpt4technicalreport} , Qwen \cite{qwen2025qwen25technicalreport} and DeepSeek-V3 \cite{deepseekai2025deepseekr1incentivizingreasoningcapability} have gained significant popularity in AI field due to their powerful language generation capabilities and versatile applications in scenarios such as chatbots \cite{openaichatgpt}, copilots \cite{microsoftcopilot} and code generation \cite{chen2021evaluating,zheng2023codegeex}. Due to their impressive capabilities, LLM inference services are widely deployed and serve many users. These services have various performance metrics, such as throughput, TTFT (Time to First Token), and TPOT (Time Per Output Token). Among these metrics, throughput—typically referring to the number of processed tokens per second—is one of the most critical indicators. High-throughput inference services can significantly expedite task processing and reduce hardware resource costs. Therefore, it is crucial to fully harness and utilize the hardware's capabilities to provide a high-throughput service.

LLM inference has garnered significant attention due to high costs, making its optimization highly valuable. 
Many researchers have dedicated themselves to developing high-throughput inference engines. 
Orca \cite{yu2022orca} proposed the continuous batching method to reduce the time and memory waste in static batching. \cite{kwon2023efficientmemorymanagementlarge} further improved memory utilization and throughput by introducing Flash-Attention mechanism. FastServe \cite{wu2024fastdistributedinferenceserving} improves throughput by enabling preemption at the granularity of each output token. 
These works primarily focus on optimizing the internal mechanism to enhance its overall capabilities.

Another research direction focuses on inference within heterogeneous clusters. 
These clusters, where machines have different capacities such as computing power and communication bandwidth, are widely prevalent within various companies. 
Many companies employ various AI accelerators, including NVIDIA GPUs, AMD GPUs \cite{amdGpus}, and NPUs (Neural Processing Units) \cite{xu2024fastondevicellminference}. 
Additionally, even within the same type of accelerator, there are many different models. 
For instance, GPUs include models like A800 and V100. 
Given the high cost of AI accelerators, the optimal strategy for these companies is to efficiently utilize their existing AI accelerators for LLM deployment. Therefore, exploring ways to provide high-throughput inference on heterogeneous clusters is highly valuable.

Unlike homogeneous clusters, heterogeneous clusters present two main opportunities for throughput improvement. First, since each machine has different resources such as computational capacity, memory, and bandwidth, the deployment configuration for each machine must be customized. An optimal configuration can significantly enhance system throughput. Second, once deployment is complete, the cluster will contain multiple inference instances, each with varying processing speeds for handling inference requests. When requests are sent to the server, they should be scheduled to different instances for processing. In homogeneous clusters, most inference engines schedule requests to instances in a round-robin fashion \cite{tensorflowserving,triton,deepspeedfastgen}, as all instances have the same computational power and memory capacity. However, in heterogeneous clusters, a round-robin algorithm can lead to unbalanced workloads and decreased system throughput, as less powerful instances become bottlenecks. Therefore, it is necessary to schedule requests while fully considering the capacities of instances and the nature of requests. In conclusion, it is crucial to explore LLM serving on heterogeneous clusters. By carefully designing deployment configurations and request scheduling strategies, we can maximize the use of heterogeneous cluster resources and achieve maximum throughput.

\begin{figure*}[htbp]
	\centering
	\includegraphics[width=\linewidth]{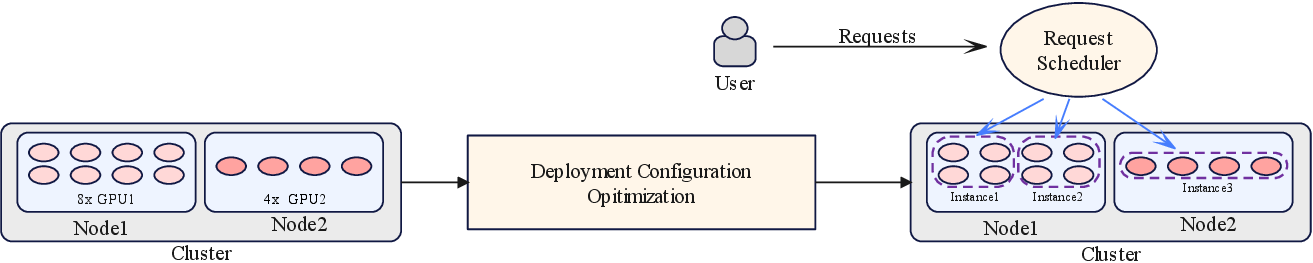}
			\vspace{-0.15in}
	\caption{An illustration of the system architecture. First, we perform deployment configuration optimization to identify the optimal deployment setup. Next, a runtime request scheduler is employed to distribute requests among instances.}
	\label{fig:arch}
\end{figure*}

In this paper, we present a novel system to achieve high-throughput LLM inference on heterogeneous clusters. 
The system architecture is illustrated in \autoref{fig:arch}.
The primary contributions can be summarized in three key aspects:\\
\textbf{Deployment Configuration}: We introduce a method to evaluate system throughput across various deployment configurations. This approach requires only lightweight profiling while avoiding resource-intensive throughput benchmarks.\\
\textbf{Request Scheduling}: For runtime scenarios, we propose a scheduler that accounts for both instance computational capacity and memory usage to efficiently distribute inference requests across different instances.\\
\textbf{Empirical Study}: We conducted two experiments on a 2-machine heterogeneous cluster. The results demonstrate that our proposed scheduling algorithm achieves a throughput improvement of 122.5\% and 33.6\%, respectively.

		\vspace{-0.2in}
		\vspace{-0.2in}
		



\section{Preliminary}
LLM inference follows an autoregressive fashion. Specifically, the user's input prompt is sent to the LLM, which, after forward propagation through the model, predicts the most likely token to appear next. This newly generated token is then concatenated with the previous input and re-entered into the model to predict the next token. This process of concatenating the new output token with the previous input and generating the next token is repeated until a special token (EOS) appears or the length of the output tokens exceeds a preset threshold. This autoregressive fashion results in an indeterminate output length and inference time for LLM inference.

\begin{figure}[htbp]
	\centering
	\includegraphics[width=0.8\linewidth]{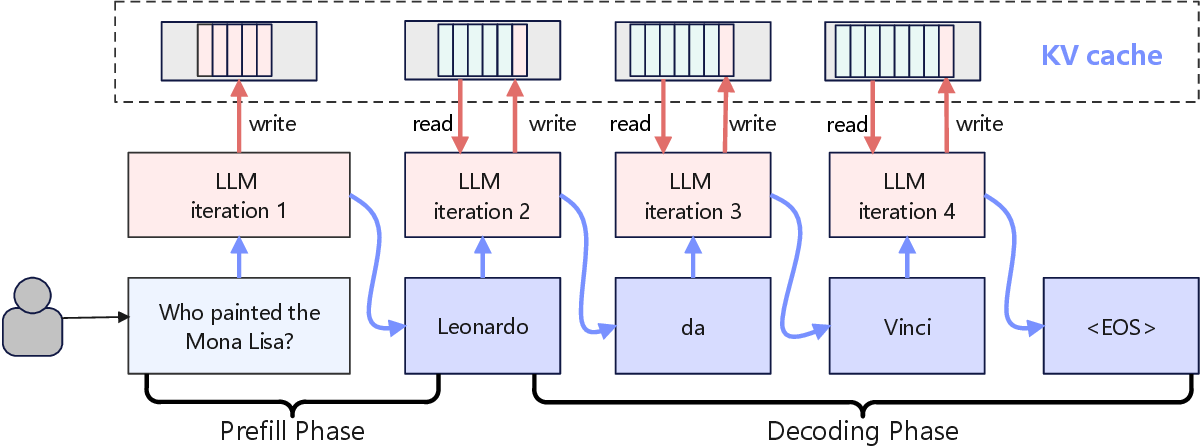}
	\vspace{-0.15in}
	\caption{An illustration of the prefill phase and decoding phase.}
	\label{fig:twoPhaseAndKVcache}
\end{figure}

Another feature of LLM inference is the two distinct stages.
The prefill phase involves taking the user's prompt as input for forward propagation and filling the KV cache. 
This process is computation-bound, with the duration correlating to the input length. The decoding phase involves continuously concatenating output tokens to the input, reading KV cache for computation, generating new token and then writing new KV cache. During this process, the frequent reading and writing operation of KV cache make it memory-bandwidth-bound. These two stages are illustrated by \autoref{fig:twoPhaseAndKVcache}.

Many technique are used in LLM inference for throughput improvement. Batching, which refers to the inference engine handling multiple requests simultaneously, is a crucial technique for improving throughput and reducing request latency. One of the simplest approaches is static batching, which involves grouping multiple requests into a batch that is processed together, returning the results only after all requests in the batch have been processed. The drawback of static batching is its lack of flexibility and the extensive padding operations it requires, which waste memory and computational resources. To address this issue, continuous batching was introduced \cite{yu2022orca}. This technique allows for the re-evaluation and selection of new batches after each token generation is completed. 
State-of-the-art inference engines like vllm, have already integrated this technology.


	\vspace{-0.2in}
	\vspace{-0.2in}

\section{Deployment Configuration optimization}\label{deployment_config_opt_section}

In this section, we focus on identifying the optimal deployment configuration for a heterogeneous cluster. 
The cluster consists of multiple machines, each equipped with multiple AI accelerators. 
While every machine contains only one type of AI accelerator, different machines may have different types of accelerators, an assumption that aligns closely with most real-world scenarios. In this study, we specifically consider the case where each instance is deployed on a single machine, requiring only tensor parallelism and eliminating the need for pipeline parallelism. 
Exploring scenarios where instances span multiple machines is left for future work.

To determine the degree of tensor parallelism for each machine, we must consider the trade-off between instance capacity and the number of instances. A higher degree of tensor parallelism results in fewer deployed instances, but each instance gains more memory and computational power, leading to faster processing speeds. Conversely, a lower degree of tensor parallelism allows for more instances, but with reduced processing speed per instance. Given this trade-off, the effectiveness of a deployment configuration must be carefully evaluated. While a straightforward approach is to test all deployment configurations by running throughput benchmarks, the cost of this method may be unacceptable.

The following proposed method can assist in finding an optimal deployment configuration through a lightweight profiling process.

\subsection{Problem Formulation}\label{deployment_config_problem_formulation}
We consider a heterogeneous cluster that contains $n$ machines, numberd from $1$ to $n$. We are going to deploy a LLM on this cluster, whose layer number is $l$ and number of hidden dimensions is $h$ and the number of parameters is $\Phi$. Let's say machine $i$ has $u_i$ AI accelerators, and each accelerator's memory capacity as $d_i$.

When deploying inference services, each machine $i$ is devided to $p_i$ instances, each with $t_j$ accelerators, satisfying $p_i\cdot t_i=u_i$. Our first objective is to decide the degree of tensor parallelism  $t_i$ for each machine.

To deploy inference services, each machine $i$ is divided into $p_i$ instances, with each instance utilizing $t_i$ accelerators, such that $p_i\cdot t_i=u_i$. The first task is to determine the degree of tensor parallelism $t_i$ for each machine.

\subsubsection{Memory constraints.} At first, a valid deployment configuration must ensure that each instance have sufficient memory for loading model and storing KV cache for at least one request, so that it has the ability to process inference requests. For an instance $s$, its available memory for storing requests' KV cache can be expressed as:
\begin{equation}\label{eq_available_mem}
	KVSize(s) = t_i \cdot d_i \cdot \phi_{usage} - \delta_{engine} - \Phi\cdot b^{bytes},
\end{equation}
where $\phi_{usage}$ denotes the proportion of memory utilized by the inference engine, while $\delta_{engine}$ refers to the amount of memory unusable by the engine due to some certain static resources (such as compiled computational graph), $b^{bytes}$ denotes the memory size that each model parameter costs, $\Phi\cdot b^{bytes}$ denotes the total memory needed for loading the model. Note that the KV cache's memory cost is strongly related to the input sequence length and generation text. Denote the maximal input length and the maximal output length of inference requests as $I_{max}$ and $O_{max}$, respectively. Then the maximal KV cache size for a single request can be $2\cdot l \cdot h\cdot(I_{max}+O_{max})\cdot b^{byte}$. Thus, the memory constraint is defined by the following equation:
\begin{equation}\label{mem_constraint}
	KVSize(s) \geq  (I_{max}+O_{max})\cdot b^{byte}, \forall i\in[1, n].
\end{equation}

\subsubsection{Instance Processing time.} 
Before discussing the overall throughput of the system, an essential step is to model the inference time cost of each instance. In this paper, we use the conclusion in \cite{cheng2024slice}, which finds that when batch size is fixed, the duration of the prefill phase is linearly related to the batch size and batch input length, while the duration of each iteration in the decode phase is linearly related to the batch size and cached length. Specifically, for instance $s$, we denote the longest input length and the longest generation text length in a processing batch $\mathcal{B}$ as $I_{\mathcal{B}}$ and $O_{\mathcal{B}}$, respectively, we have:
\begin{equation}\label{eq_prefill}
	T_{prefill}(s,\mathcal{B}) \approx p_1^s \cdot b \cdot I_{\mathcal{B}} + p_2^s \cdot b + p_3^s \cdot I_{\mathcal{B}} + p_4^s,
\end{equation}
\begin{equation}\label{eq_decode}
	\begin{aligned}
		T_{decode}(s,\mathcal{B}) &\approx \sum_{k=1}^{O_{\mathcal{B}}}\tau_{decode} (I_{\mathcal{B}} + k, b) \\
		&=\sum_{k=1}^{O_{{\mathcal{B}}}}[p_5^s \cdot b \cdot (I_{\mathcal{B}} + k) + p_6^s \cdot b + p_7^s \cdot (I_{\mathcal{B}} + k) + p_8^s],
	\end{aligned}
\end{equation}
where $b$ is the batch size of batch $\mathcal{B}$, $\tau_{decode} (I_{\mathcal{B}} + k, b)$ denotes the time of generating $k$-th token for each request, where $L_i + k$ denotes the KV cache length of each request, and $p_1^s$ to $p_8^s$ are constants and can be fitted.

For each instance, perform a lightweight profiling process to fit $p_1^s$ to $p_8^s$. The detailed steps are as follows: First, we define a set of batch sizes (e.g., 2, 4, 8, ...). For each batch size, we fix the value and sample several inference request batches, recording the input length, output length, prefill time, and decode time for each batch. The collected data is then substituted into \autoref{eq_prefill} and \autoref{eq_decode}, where parameter fitting is conducted using the least squares method to derive the values of $p_1^s$ to $p_8^s$. Note that all instances on a single machine share the same tensor parallelism degree, making it reasonable to assume they possess identical capabilities. Consequently, instances on the same machine can share the same fitted parameters.

\subsection{Optimal Deployment Configuration Searching Algorithm}
Note that 
\[
\rm throughput = \frac{\rm the\; number\; of\; tokens}{\rm processing\; time\; cost}.
\]

When evaluating the throughput of the system, we sample some requests from the dataset so that we can get a series of input length and output length. Assume that there are $q$ requests, with input length $I_r$ and output length $O_r$, where $r\in[1, q]$. For a specific deployment configuration $t_i$, we evaluate a single instance's throughput for processing these requests, denoted as $TP_s$, and the system's throughput can be calculated as $TP_{system} = \frac{TP_s \cdot u_i}{t_i}$. To calculate $TP_s$, we firstly try to group them into as big batches as possible, and then calculate each batch's processing time with \autoref{eq_prefill} and \autoref{eq_decode}. The details are shown in \autoref{algo_sys_throughput}. 

\begin{algorithm} 
	\SetKwData{currBatch}{currBatch}\SetKwData{totalTime}{totalTime}
	\SetKwData{reqNum}{batchedReqNum}\SetKwData{startIdx}{begin}
	\SetKwData{maxOutputLen}{maxO}\SetKwData{inputLenSum}{iSum}
	\SetKwData{kv}{kv}\SetKwData{tokenNum}{tokenNum}
	\SetKwFunction{calcAvailableKVSize}{calcAvailableKVSize}
	\SetKwFunction{sum}{sum}\SetKwFunction{max}{max}\SetKwFunction{calcKV}{calcKVSize}
	\SetKwFunction{calcBatchProcessingTime}{calcBatchProcessingTime}
	\SetKwInOut{Input}{input}\SetKwInOut{Output}{output}
	
	\Input{$t_i$, $u_i$, requests' input length $I_r$ and output length $O_r$, where $r\in[1, q]$} 
	\Output{System's estimated throughput under the deployment configuration}
	\BlankLine 
	$KVSize(s)\leftarrow $ \calcAvailableKVSize() \\
	\textcolor{blue}{// Group requests into batches, and sum their processing time}\\
	\totalTime $\leftarrow 0$, \reqNum $\leftarrow 0$
	
	\Repeat{\reqNum $> q$}{
		\currBatch $\leftarrow$ $[\;]$, \startIdx = \reqNum $+ 1$\\
		\For(){$r \leftarrow$\startIdx \KwTo $q$}{
			\inputLenSum $\leftarrow$ \sum($I_{begin}, I_{begin + 1}, ..., I_r$)\\
			\maxOutputLen $\leftarrow$ \max($O_{begin}, O_{begin +1}, ..., O_r$)\\
			\kv $\leftarrow$ \calcKV(\inputLenSum) $+ (r-$ \startIdx $+1)\cdot$ \calcKV(\maxOutputLen)\\
			\If{\kv $<= KVSize(s)$}{
				\currBatch $\leftarrow [begin, ..., r]$, \reqNum $\leftarrow$ \reqNum $+ 1$
			}\Else{
			    break
			}
		}
		\totalTime $\leftarrow$ \totalTime $+$ \calcBatchProcessingTime(\currBatch)
	}
	\tokenNum $\leftarrow$ \sum($I_1,...,I_q, O_1,...,O_q$)\\
	\Return{$t_i \cdot($  \tokenNum $/$ \totalTime$)$}
	
	\caption{System Throughput Estimation}
	\label{algo_sys_throughput}
\end{algorithm}

Using \autoref{algo_sys_throughput}, we can easily evaluate system throughput for all possible deployment configurations and find the optimal one. It is worth noting that the estimation is not accurate. As the state-of-the-art inference engines support advanced features like continuous-batching, which can dynamically group requests into batches and significantly improve throughput. However, our method is still meaningful as the ranking of deployment configurations in terms of throughput remains consistent between the actual results and the estimates, as shown in \autoref{sec_expr_opt_deploy}.

\section{Runtime Request Scheduling among instances}
In section \ref{deployment_config_opt_section}, we obtained an optimal deployment configuration for our heterogeneous cluster, which can be used to deploy LLM serving instances. Now, we move to the runtime scenario: with multiple running instances on different machines, each possessing varying processing capacities, how can we schedule requests among these instances to achieve maximum throughput?

Existing schedulers distribute requests to multiple LLM instances using the round-robin policy, which leads to imbalanced workloads due to the varying capacities of the instances. To solve this, we proposed a scheduler that considers both computational power and the memory usage of instances. 



\subsection{Scheduling Algorithm}

\begin{figure}
	\centering
	\includegraphics[width=\linewidth]{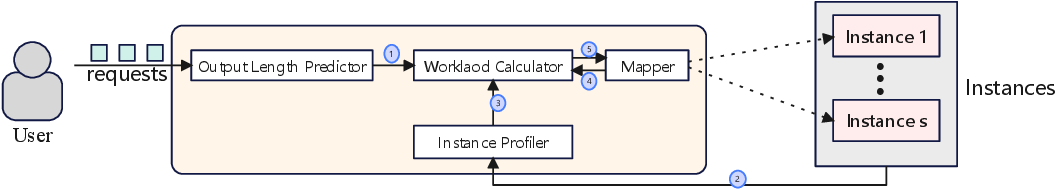}
	\caption{An illustration of the scheduler framework, in which (1) represents requests information, (2) and (3) means instances' fitted parameter, (4) means history mappings, and (5) means calculated workloads.}
	\label{fig:scheduler}
\end{figure}

To get the maximum throughput, we proposed a fast real-time scheduler. The framework and core components are illustrated in \autoref{fig:scheduler}. The scheduler contains the following core components: an instance profiler, an output length predictor, an workload calculator, and a mapper.

	\vspace{-0.15in}
	\vspace{-0.15in}

\textbf{Instance profiler}. The instance profiler works before the real requests processing, it collects running time data to fit $p_1^s$ to $p_8^s$ for instances in the cluster. 

\textbf{Output length predictor}. The output length predictor is responsible for predicting requests' output length. There are many different ways to do this, for example, we can simply predict output length based on the distribution of the dataset. Or we can integrate some existing predictors like S3 \cite{jin2023s} and Ref \cite{zheng2023response}.

\textbf{Workload Calculator}. When a request $r$ is sent to the scheduler, the workload  calculator is responsible for calculating workload for every mapping pair $(r, s)$. In this article, we aim to independently calculate the contribution of request $r$ to the workload of a specific instance $s$. Our idea is as follows: imagine that multiple identical requests $r$ are sent to instance $s$ until its KV cache size can no longer accommodate more requests, resulting in an ideal batch size $b_r^s$. We then use \autoref{eq_prefill} and \autoref{eq_decode} to calculate the processing time for these $b_r^s$ identical requests. Finally, we obtain a processing time $T_r^s$, calculated as follows:
\begin{equation}
	b_r^s = \frac{KVTotal_{s}}{KVSize(r)} = \frac{KVTotal_{s}}{2 \cdot (I_r + O_r) \cdot l \cdot h},
\end{equation}
\begin{equation}
	T_r^s = \frac{T_{prefill}(s, \mathcal{B}_r) + T_{decode}(s, \mathcal{B}_r)}{b_r^s},
\end{equation}
where $\mathcal{B}_r$ is the batch that contains $b_s^r$ identical request $r$. 

However, we cannot directly use $T_r^s$ as workload, this is because sometimes some instances' KV cache are not fully used, and their actual batch size may be much smaller than $b_r^s$. Thus, considering the KV cache utilization, we define the workload as:
\begin{equation}\label{eq_workload}
	w_r^s = T_r^s \cdot e^{\theta \cdot kvusage(s)},
\end{equation}
where $\theta > 0$ is a constant, and $kvusage(s)$ means the proportion of instance $s$'s KV cache that has already been utilized. Assume that for the requests sent to instance $s$
, the sum of the input lengths and the sum of the predicted output lengths for the unfinished requests are $I_{sum}^{running}$ and $\tilde{O}_{sum}^{running}$, respectively. Then we have:
\begin{equation}
	kvusage(s) = \frac{2\cdot(I_{sum}^{running} + \tilde{O}_{sum}^{running})\cdot l \cdot h\cdot b^{byte}}{KVSize(s)}.
\end{equation}
Note that $kvusage(s)$ can be larger than 1, as we need to need to account for the scenario where too many requests are sent to instance $s$, causing some requests to wait for processing.

Our proposed workload calculation method \autoref{eq_workload} takes into account both the computational capacity of the instances and the load on their KV cache. When the request arrival rate is moderate and none of the instances' KV caches are overloaded, the term $T_r^s$
in the formula holds greater weight, favoring the allocation of requests to instances with stronger computational capabilities. However, in scenarios such as a lot of requests simultaneously arriving at the server, the $kvusage(s)$ of the instances increases rapidly. Consequently, the second term in the formula also rises significantly, and the request scheduling primarily shifts to being based on the memory capacity of each instance.

\textbf{Mapper}. The mapper is the component responsible for deciding the mapping from requests to instances. It receives the calculator's results and, based on the current workloads of the instances, distributes the request to an instance in such a way as to  minimize $max(\rm workloads\; of \; instances)$. Meanwhile, the mapper maintains a list that records the workloads of all instances. When the mapper distributes a request $r$ to an instance $s$, it updates the workload of that instance $s$ by adding $w_r^s$. Additionally, the mapper adds a hook function that, upon the completion of the request $r$, reduces the instance $s$'s workload by $w_r^s$. This hook function prevents the accumulation of errors that could undermine the algorithm's effectiveness. This algorithm is illustrated in \autoref{algo_mapper}

	\vspace{-0.15in}
\begin{algorithm} 
	\SetKwData{currBatch}{currBatch}\SetKwData{totalTime}{totalTime}
	\SetKwData{startIdx}{begin}
	\SetKwData{maxOutputLen}{maxO}\SetKwData{inputLenSum}{iSum}
	\SetKwData{kv}{kv}\SetKwData{tokenNum}{tokenNum}
	\SetKwData{reqNum}{r}
	\SetKwData{instWorklaods}{instLoads}\SetKwData{instRunningReqLength}{instRunningReqLen}
	\SetKwData{minGlobalWorkload}{minGlobalLoad}\SetKwData{currWorkload}{tmp}
	\SetKwData{choice}{c}
	\SetKwFunction{calcAvailableKVSize}{calcAvailableKVSize}
	\SetKwFunction{sum}{sum}\SetKwFunction{max}{max}
	\SetKwFunction{calcBatchProcessingTime}{calcBatchProcessingTime}
	\SetKwFunction{calcT}{calcT}\SetKwFunction{calcKV}{calcKVSize}
	\SetKwFunction{calcWorkload}{calcWorkload}
	\SetKwFunction{sendRequest}{sendRequest}
	\SetKwFunction{addHook}{addHook}
	\SetKwInOut{Input}{input}\SetKwInOut{Output}{output}
	
	\Input{the number of the requests $N_{req}$, each with input length $I_r$ and predicted output length $\tilde{O}_r$, the number of instances $N_{inst}$.} 
	\Output{the scheduling results}
	\BlankLine 
	$KVSize(s)\leftarrow $ \calcAvailableKVSize($s$), for $s\in[1, N_{inst}]$ \\
	\instWorklaods $\leftarrow [0, ..., 0]$, \instRunningReqLength $\leftarrow [0, ..., 0]$\\
	
	\For(\textcolor{blue}{// Decide mapping for each request $r$}){\reqNum $\leftarrow 1$ \KwTo $N_{req}$} {
		\For() {$s \leftarrow1$ \KwTo $N_{inst}$} {
			$T_r^s \leftarrow$ \calcT($I_r$, $\tilde{O}_r$, $KVSize(s)$)\\
			$kvusage(s) \leftarrow$ \calcKV(\instRunningReqLength$[s]$) $/ KVsize(s)$\\
			$w_r^s\leftarrow $ \calcWorkload($T_r^s, kvusage(s)$)
		}
		\minGlobalWorkload $\leftarrow 0$, \choice $\leftarrow 1$ \textcolor{blue}{// $c$ means final chosen instance}\\
		\For() {$s \leftarrow1$ \KwTo $N_{inst}$} {
			\currWorkload $\leftarrow $ \max( \instWorklaods$[1], ...,$ \instWorklaods$[s] + w_r^s,...,$  \instWorklaods$[N_{inst}]$)\\
			\If{\minGlobalWorkload $>$ \currWorkload}{
				\minGlobalWorkload $\leftarrow$ \currWorkload\\
				\choice $\leftarrow s$ 
			}
		}
		\sendRequest($r$, \choice)\textcolor{blue}{// Send request $r$ to instance \choice}\\
		\instWorklaods$[$\choice$]$ $\leftarrow$ \instWorklaods$[$\choice$]$ $+ w_r^{choice}$\\
		\instRunningReqLength$[$\choice$]$  $\leftarrow$ \instRunningReqLength$[$\choice$]$ $+ I_r + \tilde{O}_r$\\
		\addHook(when $r$ is completed, \instWorklaods$[$\choice$]$ $\leftarrow$ \instWorklaods$[$\choice$]$ $- w_r^{c}$)\\
		\addHook(when $r$ is completed, \instRunningReqLength$[$\choice$]$  $\leftarrow$ \instRunningReqLength$[$\choice$]$ $- I_r - \tilde{O}_r$)
	}
	
	\caption{Request Scheduling among instances}
	\label{algo_mapper}
\end{algorithm}
	\vspace{-0.15in}

\section{Evaluation}
In this section, we conduct three experiments to show the effectiveness of our algorithm. Firstly, we show that the the proposed method can find the best deployment configuration for a single machine. Secondly, we show that the scheduling algorithm can effectively reduces workload imbalance among instances. In the third experiment, we evaluate the throughput improvement in a 2-machine heterogeneous cluster.

\subsection{Optimal Deployment Configuration Searching}\label{sec_expr_opt_deploy}
The details of the experiment is illustrated as follows:

\textbf{Testbed}. The experiment was conducted on a machine equipped with 8 Nvidia Tesla V100-SXM2-32GB GPUs(connected over PCIe 3.0 x16), 2 Intel Xeon Gold 6132 CPUs, and 516 GB host memory.

\textbf{Models}. We deployed the Meta-Llama-3-8B model on this machine, and load the model weights as FP16 format, which means each parameter cost 2 bytes of GPU memory. Under the memory constraint, we may choose to set $t_i$ as 1, 2, 4, or 8, corresponding to 8, 4, 2, and 1 instances, respectively.

\textbf{Inference Engine}. We use vllm 0.6.0 as inference engine, which is one of state of the art engines in the community. It supplies advanced features like Paged Attention and continuous-batching. When deploying the model, the gpu memory utilization fraction is set to be 0.9, and vllm engine takes 1-3 GB memory for cuda graphs per GPU, thus we set $\phi_{usage}=0.9$ and $\delta_{engine}=2 GB$ in \autoref{eq_available_mem}. When instances are deployed, we conduct some profiling to sample some latency data, and then use $curve_fit$ function from scipy library \cite{virtanen2020scipy}  to fit constants $p_1^s$ to $p_8^s$ in \autoref{eq_prefill} and \autoref{eq_decode}.

\textbf{Overall System Throughput Metrics}. We randomly sample prompts from ShareGPT\_Vicuna\_unfiltered dataset, which originates from ShareGPT conversations and is widely used by community. To thoroughly evaluate the reasoning capabilities of each instance, we set the request rate to be $inf$, which means that we send all requests to the request pool at once and distribute them to each instance without any delay. 

\begin{figure*}[htbp]
	\centering
	\includegraphics[width=\linewidth]{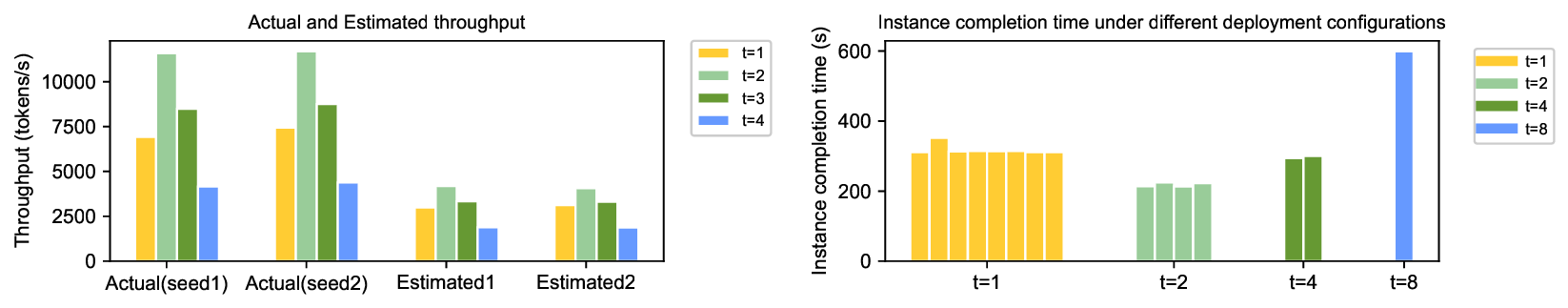}
	\caption{System throughput and instance completion time distribution under different deployment configurations. The first subgraph shows the throughput under two different random seeds and two estimated results. The ranking of deployment configurations in terms of throughput remains consistent between the actual results and the estimates. The second subgraph illustrates the completion times of each instance in the system under the first random seed. With the guarantee of our designed request scheduling mechanism, there is no workload imbalance in the system. }
	\label{fig:config_opt}
\end{figure*}

Meantime, to prevent workload imbalance from hindering a thorough evaluation of the system's throughput capabilities, we designed the following request scheduling mechanism: First, we randomly sampled 1,000 examples from the dataset and duplicated these samples 8 times. For example, the original requests $[r1, r2, r3,…]$ are duplicated into $[r1^{(1)},…, r1^{(8)}, r2^{(1)},…, r2^{(8)},…]$. Next, we distributed these requests evenly across all instances in a round-robin fashion. This ensures that when there are multiple instances in the system, each instance has an identical workload. For example, with 4 instances, each instance processes a request list like $[r1, r1, r2,r2,…]$. Additionally, for any given $t_i$, the entire system maintains an identical load, with all instances collectively handling the same 8,000 requests. As shown in the second subgraph in \autoref{fig:config_opt}, when there are multiple instances, their completion times, which refers to the time taken to complete all inferences, are almost identical, meaning that there is no workload imbalance among the instances in the system, ensuring the accuracy of the throughput metrics. 

We conducted two experiments using different random seeds, which resulted in the system receiving different request lists from the dataset. 

\textbf{System Throughput Evaluation}. When using \autoref{algo_sys_throughput} to evaluate the system throughput, we need to input a series of requests' input lengths and output lengths. Therefore, we randomly sampled requests twice from the dataset, with each sample containing 200 requests, and then estimated the system's throughput based on the input length and output length of these requests.

	\vspace{-0.1in}
	\vspace{-0.15in}

\textbf{Result}. As shown in \autoref{fig:config_opt}, we can see that different deployment configurations leads to big difference in system's overall throughput. And the throughput remains consistent across the two different random seeds. The configuration resulting in throughput from highest to lowest is $t=2$, $t=4$, $t=1$, $t=8$. Meanwhile, we observed that the estimated throughput differs significantly from the actual throughput. This discrepancy is reasonable because, in our algorithm, when grouping multiple requests into a batch, we consider a static-batching model, whereas vllm integrates state-of-the-art methods such as continuous batching, which can significantly enhance throughput. However, we also observed that our algorithm maintains order-preserving consistency with the actual results. Based on the two different sampled request lists, the estimated results for throughput both show a consistent pattern of decreasing in the order of $t=2$, $t=8$, $t=1$, and $t=8$, which aligns with the actual sequence.

\subsection{The impact of the request scheduler on workload balancing}
In this part, we aim to demonstrate that our scheduler can effectively alleviate workload imbalances among instances.

\textbf{Experiment Setting}. We still conduct this experiment on the 8x V100 machine. This time, we deploy two Meta-Llama-3-8B model instances on it, one with $t=4$ and another with $t=1$. Obviously, these two instances have significantly different capabilities in handling requests, with the first instance achieving much higher throughput. To evaluate the system throughput, we randomly sample 4000 requests from the dataset, and send them to the scheduler under different request rates. The request rates are 8 req/s, 16 req/s, 24 req/s and inf. For output length prediction, we employed a simple predictor. We sampled a subset of requests from the dataset to calculate the mean and standard deviation of their output lengths. Then, we used the $numpy.random.normal$ method from NumPy to generate random values as predictions.

\textbf{Scheduling Strategies}. We evaluated the throughput under the following scheduling strategies: (1) Round Robin scheduling (\textbf{RR}), where requests are distributed to two instances with equal probability, (2) Single instance scheduling (\textbf{SI}), meaning dispatching all requests to the first instance only, (3) Memory based scheduling (\textbf{MB}), which is our proposed scheduler with $\theta$ set to 2, but always set $T_r^s = 1$, meaning that this scheduler only considers the memory usages of instances to schedule requests, (4) Our scheduler (\textbf{OS}), with $\theta$ set to 2, (5) Weighted Round Robin scheduling \textbf{WRR}, where the weights for the first and second instances are set to be 4 and 1 respectively, meaning that requests are sent to the first instance with a probability four times higher than that of the second instance.

		\vspace{-0.15in}
\begin{figure*}[htbp]
	\centering
	\includegraphics[width=\linewidth]{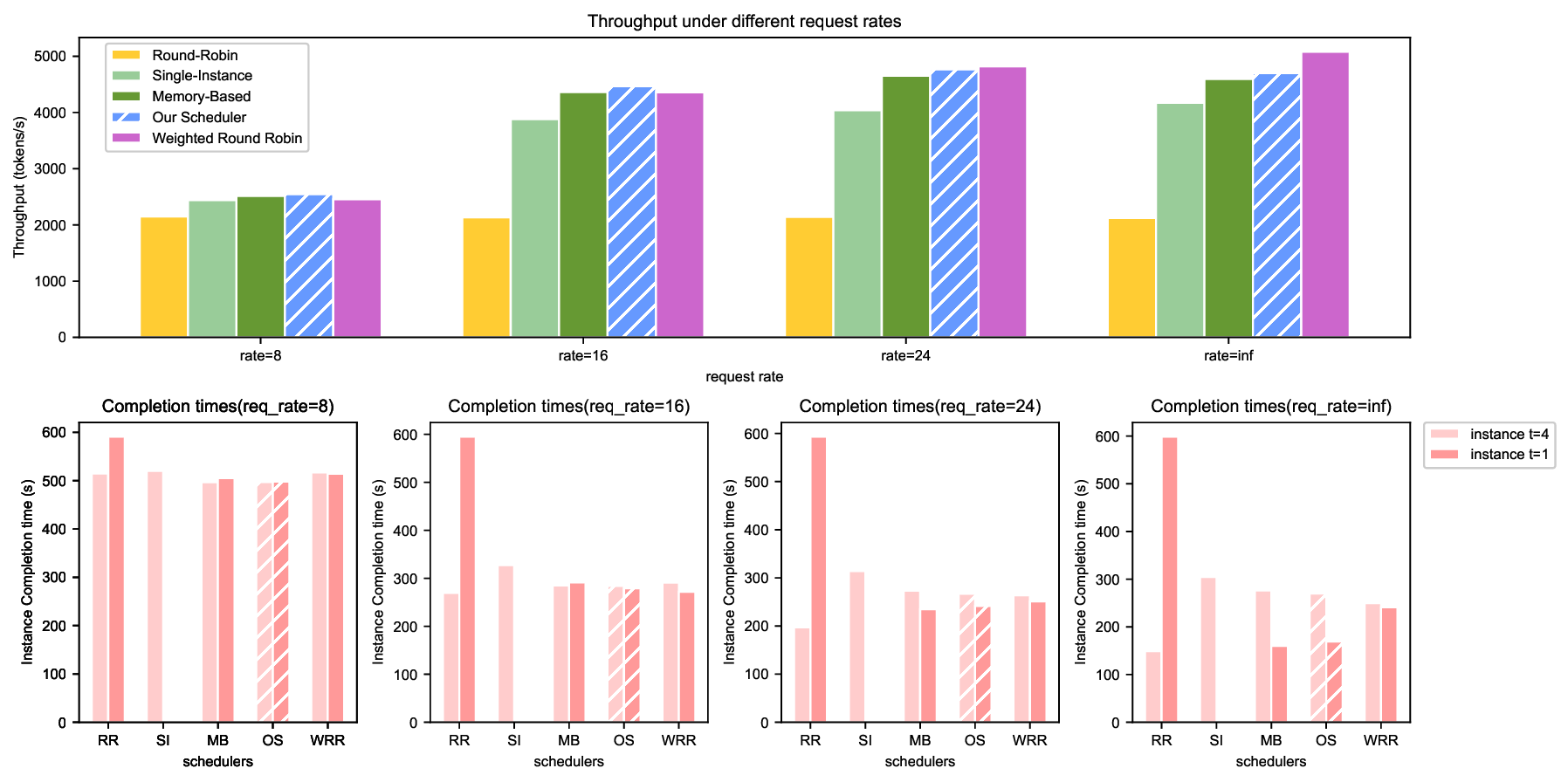}

	\caption{The throughput and instance completion time distribution for a system with two instances, one with t=4 and the other with t=1, demonstrate that the Round Robin scheduling strategy leads to extreme workload imbalance between instances, which becomes a bottleneck for system throughput. In contrast, our scheduler significantly reduces this imbalance.}
	\label{fig:result_t2t4}
\end{figure*}
		\vspace{-0.15in}

\textbf{Result}. As shown in \autoref{fig:result_t2t4}, the system throughput varies significantly under different scheduling strategies. We observe the following:

Firstly, the Round Robin strategy delivers the lowest throughput, even lower than the strategy of dispatching all requests to the stronger instance. This is because the two instances differ greatly in computational power and KV cache size, causing the weaker instance (the one with $t=1$) to become the system's throughput bottleneck. 
A significant difference exists in completion time between the two instances, which demonstrates the imbalance in workload distribution.

Secondly, when the request rate is high (request rate = 24 and infinity), the weighted Round Robin strategy achieves the highest throughput. However, this experiment is rather specific: since the first instance uses over 4 times the GPU resources, its weight can naturally be set to four times that of the other instance. Nevertheless, in large-scale heterogeneous clusters with multiple GPU models, it becomes challenging to determine appropriate weights for each instance.

Thirdly, our proposed scheduling algorithm also performs exceptionally well in terms of throughput. At request rates of 8 and 16, it achieves the highest throughput. At a request rate of 24, our scheduler achieves a 122.5\% increase in throughput compared to the round-robin scheduling strategy. Additionally, in terms of scheduling instance completion times, the difference in completion times between the two instances is much smaller than with the Round Robin strategy, indicating a significant improvement in workload balance among instances.

Finally, the memory-based scheduling strategy, which always sets $T_r^s = 1$, performs slightly worse in throughput compared to our scheduling algorithm. This indicates that our approach to calculating the workload of individual requests brings a certain level of performance gain.

\subsection{Throughput Estimation in Multi-Machine Heterogeneous Clusters}
\begin{figure*}[htbp]
	\centering
	\includegraphics[width=\linewidth]{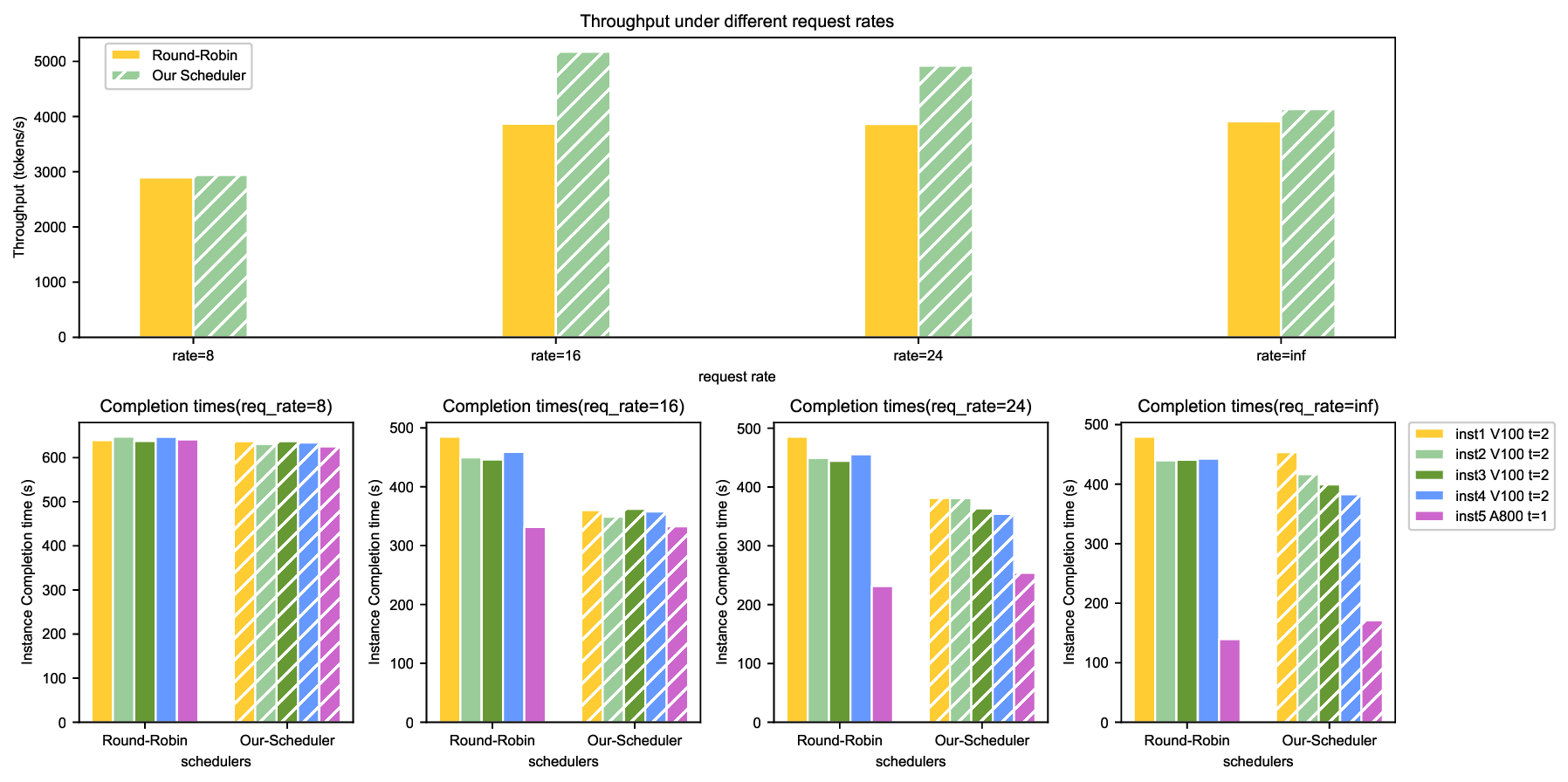}
	\caption{The test results on a heterogeneous cluster composed of two machines. One machine hosts 4 instances of V100 with a tensor parallelism degree of $t=2$, while the other machine hosts 1 instance of A800 with a tensor parallelism degree of $t=1$.}
	\label{fig:result_Vt2x4_At1x1}
\end{figure*}

In this part, we conducted tests using two machines. One of the machines was the same as used in previous experiments, equipped with 8x V100 GPUs. The other machine had 1x NVIDIA A800 80GB PCIe GPU and two Intel(R) Xeon(R) Gold 6240 CPUs. On the first machine, we deployed four DeepSeek-R1-Distill-Qwen-14B models, with each instance having a tensor parallelism degree of $t=2$. On the second machine, we deployed a single model with a tensor parallelism degree of $t=1$. 
As illustrated in \autoref{fig:result_Vt2x4_At1x1}, our algorithm achieves higher throughput across various request rates. At a request rate of 16, the system throughput increased by 33.6\%.

		\vspace{-0.15in}
		\vspace{-0.15in}

\section{Conclusion}
In this paper, we search the optimal deployment configuration based on modeling instance memory and batch processing time, which only requires lightweight profiling data rather than heavy throughput benchmarks. 
Besides, a novel request scheduler is proposed to address load imbalance. 
Comprehensive experiments on heterogeneous clusters demonstrate $\geq$33.6\% improvement in throughput.

%
%
%
\bibliographystyle{splncs04}
\bibliography{mybibliography}
\end{document}